\newcommand{\Tr}{\mathop{\mathrm{Tr}}\nolimits}
\newcommand{\tr}{\mathop{\mathrm{tr}}\nolimits}

\newcommand{\opx}{\hat{\sigma}_x}
\newcommand{\opy}{\hat{\sigma}_y}
\newcommand{\opz}{\hat{\sigma}_z}
\newcommand{\identity}{\hat{\openone}}

\documentclass[osajnl,eqsecnum,twocolumn,showpacs]{revtex4}

\usepackage{amsfonts,amssymb,amsbsy,bm,subeqnarray,graphicx}

\begin{document}

\title{Mutually unbiased bases and discrete Wigner functions}

\author{Gunnar Bj\"{o}rk}
\affiliation{School of Information and Communication Technology,
Royal Institute of Technology (KTH), Electrum 229, SE-164 40 Kista,
Sweden}

\author{Jos\'e L. Romero}
\affiliation{Departamento de F\'{\i}sica,
Universidad de Guadalajara,
44420~Guadalajara, Jalisco, Mexico}

\author{Andrei B. Klimov}
\affiliation{Departamento de F\'{\i}sica,
Universidad de Guadalajara, 44420~Guadalajara,
Jalisco, Mexico}

\author{Luis L. S\'anchez-Soto}
\affiliation{Departamento de \'{O}ptica, Facultad de F\'{\i}sica,
Universidad Complutense, 28040~Madrid, Spain}

\date{\today}

\begin{abstract}
Mutually unbiased bases and discrete Wigner functions are
closely, but not uniquely related. Such a connection becomes
more interesting when the Hilbert space has a dimension that
is a power of a prime $N=d^n$, which describes a composite
system of $n$ qudits. Hence, entanglement naturally enters
the picture. Although our results are general, we concentrate
on the simplest nontrivial example of dimension $N=8=2^3$.
It is shown that the number of fundamentally different Wigner
functions is severely limited if one simultaneously imposes
translational covariance and that the generating operators
consist of rotations around two orthogonal axes, acting on
the individual qubits only.
\end{abstract}


\maketitle

\section{Introduction}
\label{Introduction}

Quasiprobability distributions have been useful in quantum mechanics
both as visualization tools and for computational purposes. Typical
examples include the Glauber-Sudarshan~\cite{Glauber,Sudarshan},
the Husimi~\cite{Husimi}, and the Wigner function~\cite{Wigner}.
These are all functions of continuous parameters that map
quantum states onto a continuous phase space.

The growing interest in quantum information has fueled the
search for discrete phase-space counterparts of these
distributions~\cite{Vourdas}. In particular, the discrete
Wigner function has attracted a good deal of attention.
There is no unique way of defining such a function:
one family of methods maps the states of the Hilbert
space onto a phase-space grid of $2N \times 2N$ points
(implying a redundancy of information)~\cite{Hannay,Leonhardt1,Leonhardt2},
whereas it is also possible to map the states onto
a (nonredundant) $N \times N$ phase
space~\cite{Buot,Wootters0,Galetti,Cohendet,Wootters4,Gibbons}.
In this paper we shall investigate the latter type of Wigner
functions.

Even-dimensional Hilbert spaces are of special interest to
visualize the effects of quantum information processing.
The ubiquitous qubit lives in a two-dimensional Hilbert
space, and $n$ qubits hence span a $2^n$-dimensional
space. Wigner functions for two and three qubits have
found applications in providing solutions to, e. g.,
state reconstruction~\cite{Asplund2}, quantum
teleportation~\cite{Koniorczyk,Paz}, quantum
optics~\cite{Vaccaro}, quantum computing~\cite{Miquel}, and
the mean king problem~\cite{Aharonov,Englert,Aravind1,Aravind2,Durt,Hayashi}.

In several of these contexts, mutually unbiased bases (MUBs)
are also of interest~\cite{Delsarte,Wootters}. We recall that
each vector in one of these bases is an equal-magnitude
superposition of all the vectors in any of the other bases.
MUBs are central to quantum tomography and state
reconstruction~\cite{Wootters2}, but are also valuable for
quantum key distribution~\cite{Cryptography,Bechmann,Asplund,Bruss}.
It is known that when the Hilbert space dimension $N$ is a prime,
or a power of a prime, there exist exactly $N+1$ MUBs~\cite{Ivanovic}.
In this paper we shall mainly discuss the three-qubit case, for
it is the smallest space consisting of qubits where different
MUB structures, with respect to their entanglement properties,
exist~\cite{Lawrence,Romero}. More precisely, in this space
there are four different MUB structures corresponding to $s=0,
1, 2,$ and 3, respectively, where $s$ is the number of
triseparable MUBs. The fact that different structures are
possible has strong implications for the physical implementation
of quantum tomographic measurements and quantum information
protocols. It also affects the manner in which we can map
states onto a phase space while retaining some properties
that we consider ``natural" in the continuous case.

The rest of the paper is organized as follows: In
Section~\ref{Continuous Wigner} the continuous Wigner
function and specifically the concept of translational
covariance is recalled. In  Section~\ref{MUB} a method
for determining MUBs for three qubits is briefly reviewed.
The concept involves operator sets based on primitive,
one-qubit operations. In Sec.~\ref{Discrete Wigner} we
devise a construction of the Wigner function by associating
the primitive operators (that generate a certain MUB) with
translations in phase space. We show that this MUB  has
a unique structure, once we disregard the somewhat trivial
degrees of freedom related to qubit labeling  and to
association of lines or curves in phase space to states.
In Section~\ref{The (3,0,6)} we derive two different Wigner
functions from a different MUB structure, while in
Section~\ref{Negative result} we discuss the remaining two MUB
structures, having $s=0$ or $s=1$ triseparable bases. In this
case, the principle of translational invariance no longer works
in association with translations corresponding to single qubit
MUB-generating operators. Finally, we make some concluding
remarks in Section~\ref{Conclusions}.

\section{The continuous Wigner functions}
\label{Continuous Wigner}

The continuous Wigner function has three properties that
we are particularly keen on retaining when constructing
a discrete counterpart. The function is real, when
integrated along any direction in phase space it yields
a nonnegative function with unit area, i.e., a
marginal probability distribution, and finally it is
translationally covariant. The latter property can be
stated in mathematical terms as follows. Let $W(q, p)$ be
the Wigner function (expressed in terms of the position $q$
and the momentum $p$) corresponding to a density matrix
$\hat{\rho}$, and let $\hat{\rho}^\prime$ obtained from
$\rho$ by a displacement $(q_0,p_0)$ in phase space:
\begin{equation}
\hat{\rho}^\prime = \exp[i(q_0 \hat{p} - p_0 \hat{q})/\hbar] \,
\hat{\rho} \, \exp[-i(q_0 \hat{p} - p_0 \hat{q})/\hbar] ,
\end{equation}
where $\hat{q}$ and $\hat{p}$ are the position and momentum
operators, respectively. Then the Wigner function $W^\prime$
corresponding to $\hat{\rho}^\prime$  is obtained from
$W$ via the transformation
\begin{equation}
W^\prime (q, p) = W (q-q_0, p-p_0) .
\end{equation}
In other words, when the density matrix is translated, the
Wigner function follows along rigidly.

\section{MUB structures for three qubits}
\label{MUB}

Mutually unbiased bases can be constructed using a number of
methods that depend on the dimensionality of the space. The
main dividing lines are if the dimension is prime, a product
of primes, or a power of a prime, and, in the latter case, if
it is odd or even~\cite{Calderbank,Calderbank2,Chaturvedi,Pittenger,Durt2,Bandyopadhyay,Klimov,Grassl,Wocjan,Archer}.
The problem appears to be closely related to mutually orthogonal
Latin squares~\cite{Klappenecker,Wootters 5} and to the existence
of finite projective planes of certain orders~\cite{Saniga,Bengtsson}.
As stated in the Introduction, we confine our study to the case of
three qubits, that is, to an eight-dimensional Hilbert space. In this
Hilbert space there exist four MUB structures, where the word
``structure'' denotes sets of MUBs where the basis vectors are
either triseparable (i.e., factorizable in a tensor product of
three individual qubit states), separable in one qubit and one
maximally entangled two-qubit state (i.e., biseparable), or
nonseparable~\cite{Lawrence,Romero}. The four structures are
(3,0,6), (2,3,4), (1,6,2), and (0,9,0), where the labels indicate
the number of triseparable, biseparable, and nonseparable bases,
respectively. Note that all of them are of the form $(s,b,9-s-b)$,
where $b$ is the number of biseparable bases. We shall primarily
concentrate on the two first structures, since they correspond to
MUBs where at least two bases are fully separable so we can
associate theses bases with local properties of the three qubits.
This simplifies the analogy with the continuous Wigner function.

We follow the construction algorithm  worked out by Klimov,
S\'anchez-Soto, and de Guise in Ref.~\onlinecite{Klimov} to
generate, in a systematic way, a set of operators whose
eigenvectors constitute a MUB. If we take a spin-1/2 system
as our model for a qubit (described, apart from a factor
$\hbar/2$, by the Pauli operators $\hat{\bm{\sigma}}$), for
three qubits it suffices to consider tensor products of
$\identity^{(i)}, \opx^{(i)}, \opy^{(i)}, \opz^{(i)}$, where
the superscript $i = 1, 2, 3$ labels the qubit, in such a way
that the tensor product contains one operator acting on each
qubit. For example, $\opx^{(1)} \otimes \identity^{(2)} \otimes
\opy^{(3)}$ is a legitimate operator in this respect, which
we shall write from now as $\opx \identity \opy$.

We wish now to construct MUBs where two of the bases are generated
by individual qubit rotations around two orthogonal axes, which we
arbitrarily take as the $x$ and the $z$ axes. Hence, we define the
MUB using the operators $\{\opz \identity \identity, \identity \opz
\identity, \identity \identity \opz\}$ and $\{ \opx \identity
\identity, \identity \opx \identity, \identity \identity \opx\}$. It
can be shown that fundamentally, there is only a single way of
accomplishing this, namely from the construction in Table~\ref{table1}.

\begin{table}
\caption{Nine sets of operators defining a (2,3,4) MUB.}
\begin{tabular}{c c c c c c c}   \hline \hline
$\opz \identity \identity$ & $\identity \identity \opz$ & $\identity
\opz \identity$ & $\opz \identity \opz$ &
$\identity \opz \opz$ & $\opz \opz \opz$ & $\opz \opz \identity$ \\
$\opx \identity \identity$ & $\identity \opx \identity $ &
$\identity \identity \opx$ & $\opx \opx \identity$ & $
\identity \opx \opx$ & $\opx \opx \opx$ & $\opx \identity \opx$ \\
$\opy \identity \identity$ & $\identity \opx \opz$ & $\identity \opz
\opx$ & $\opy \opx \opz$ & $\identity \opy \opy$ &
$\opy \opy \opy$ & $\opy \opz \opx$ \\
$\opx \identity \opz$ & $\identity \opy \identity$ & $\opz \identity
\opy $ & $\opx \opy \opz$ & $\opz \opy \opy$ &
$\opy \opy \opx$ & $\opy \identity \opx$ \\
$\opx \opz \identity$ & $\opz \opx \opz$ & $\identity \opz \opy$ &
$\opy \opy \opz$ &
$\opz \opy \opx$ & $\opy \opx \opx$ & $\opx \identity \opy$ \\
$\opy \identity \opz$ & $\identity \opy \opz$ & $\opz \opz \opy$ &
$\opy \opy \identity$ & $\opz \opx \opx$ & $\opx \opx \opy$ &
$\opx \opz \opx$  \\
$\opx \opz \opz$ & $\opz \opy \opz$ & $\opz \opz \opx$ & $\opy \opx
\identity$ & $\identity \opx \opy$ & $\opx \opy \opx$ &
$\opy \identity \opy$ \\
$\opy \opz \opz$ & $\opz \opy \identity$ & $\opz \identity \opx$ &
$\opx \opx \opz$ & $\identity \opy \opx$ & $\opy \opx \opy$ &
$\opx \opz \opy$ \\
$\opy \opz \identity$ & $\opz \opx \identity $ & $\identity
\identity \opy$ & $\opx \opy \identity$ &
$\opz \opx \opy$ & $\opx \opy \opy$ & $\opy \opz \opy$ \\
\hline \hline
\end{tabular}
\label{table1}
\end{table}

The algorithm guarantees that the simultaneous eigenstates
of the operators in each of the nine rows give a complete
basis, and each basis is mutually unbiased to each other.
From earlier work~\cite{Lawrence,Romero}, we know that
with respect to separability, Table~\ref{table1} defines
a (2,3,4) structure.

Moreover, the table is uniquely defined by the $2 \times 3$ entries
in the three first columns of the first two rows. Indeed this is
the case, since the operators in the first two rows are related
by $\hat{O}_{r,c} = \hat{O}_{r,c-2} \hat{O}_{r,c-3}$, where $r=
1,2$ and $c = 1, \ldots , 7$ denote the row and column of the
operator, respectively, and must be taken modulo seven.
Subsequently, the other rows can be expressed as $\hat{O}_{r,c} =
\hat{O}_{2,c} \hat{O}_{1,c+r-3}$ for $2 < r \leq 9$ (where all
phase factors arising from the products have been neglected).
In what follows, we write explicitly only the first $2 \times 3$
elements of each basis to save space.

\begin{table*}
\caption{The five different sets resulting from a rearranging of the
three qubits of Table~\ref{table1} as 132, 213, 312, 321, and 231,
respectively. Only the relevant $2 \times 3$ operators are shown.}
\begin{tabular}{ccc} \hline \hline
$\opz \identity \identity$ & $\identity \opz \identity$ &
$\identity \identity \opz$ \\
$\opx \identity \identity$ & $\identity \identity \opx$ & $\identity
\opx \identity$ \quad \\ \hline \hline
\end{tabular}
\hspace*{0.25cm}
\begin{tabular}{ccc} \hline \hline
$\identity \opz \identity$ & $\identity \identity \opz$ &
$ \opz \identity \identity$ \\
$\identity \opx \identity$ & $\opx \identity \identity $ &
$\identity \identity \opx$ \\  \hline \hline
\end{tabular}
\hspace*{0.25cm}
\begin{tabular}{ccc} \hline \hline
$\identity \opz \identity$ & $\opz \identity \identity$ &
$\identity \identity \opz$ \\
$\identity \opx \identity$ & $\identity \identity \opx $ & $\opx
\identity \identity$ \\  \hline \hline
\end{tabular}
\hspace*{0.25cm}
\begin{tabular}{ccc} \hline \hline
$\identity \identity \opz$ & $ \opz \identity \identity$ &
$\identity \opz \identity$  \\
$\identity \identity \opx$ & $\identity \opx \identity$ & $\opx
\identity \identity $ \\  \hline \hline
\end{tabular}
\hspace*{0.25cm}
\begin{tabular}{ccc} \hline \hline
$\identity \identity \opz$ & $\identity \opz \identity$ &
$\opz \identity \identity$  \\
$\identity \identity \opx $ & $\opx \identity \identity$ &
$\identity \opx \identity$  \\  \hline \hline
\end{tabular}
\label{table2}
\end{table*}

We observe that if the first and second row are interchanged, the
new table defines the same structure. Our choice of axes is of
course arbitrary, we could just as well denote the $z$ axis by $x$
and vice versa. If we permute the labeling of the axes, the result
is the first set in Table~\ref{table2}. This looks different from
Table~\ref{table1}, and for fixed $x$ and $z$ axes it generates a
different MUB than Table~\ref{table1}. However, at the fundamental
level, both are the same structure. If a certain election of the
axes orientation is made, the MUBs are identical.

Another arbitrary choice we have made is the labeling
of the qubits. If the qubits are numbered 1, 2, and 3,
from left to right in Table~\ref{table1}, rearranging the
qubits (and the corresponding operators) in the orders 132,
213, 312, 321, and 231 will then result in the five sets
shown in Table~\ref{table2}. We see that, due to the simple
structure of the tables, a permutation of the axes can also
be seen as a relabeling of the qubits.

The sets in Tables~\ref{table1} and \ref{table2} are the
only three-qubit, MUB generating matrices containing in
one row the single qubit rotations around one axis, e.g.,
$\{\opz \identity \identity, \identity \opz \identity,
\identity \identity \opz\}$ (in any order), and in the
next row the single qubit rotations around an orthogonal
axis, e.g., $\{ \opx \identity \identity, \identity \opx \identity,
\identity \identity \opx\}$. Hence, with this restriction,
only one MUB structure is allowed (up to a relabeling
of the qubits). This is our first important result.

\section{A three-qubit Wigner function}
\label{Discrete Wigner}

When the Hilbert space is finite, there are several possibilities
for defining a Wigner function, as discussed in the Introduction.
We follow the approach of Wootters, Gibbons, and
Hoffman~\cite{Wootters4,Gibbons} and try to incorporate the three
features discussed for the continuous case. To generate Wigner
functions from the MUB in Table~\ref{table1}, we resort to
elementary notions of finite fields. The field $\mathbb{F}_8$ with
exactly eight elements can be seen as the set $\{0,1,\mu, \ldots
,\mu^6\}$ (which is also an additive group), where the primitive
element $\mu$ is a root of the following irreducible polynomial on
$\mathbb{Z}_{2}$ (the integers modulo 2)
\begin{equation}
\theta^{3} + \theta + 1 = 0 .
\label{Eq: Irreducible}
\end{equation}
With this arithmetic, we have
\begin{eqnarray}
& 1 = \mu^3 + \mu^5 + \mu^6 , \qquad
\mu = \mu^5 + \mu^6 , & \nonumber \\
& & \\
& \mu^2  =  \mu^3 + \mu^5 , \qquad
\mu^4 = \mu^3 + \mu^6 . & \nonumber
\label{Eq: Field element powers}
\end{eqnarray}

Note that the subset $\{\mu^3,\mu^5,\mu^6\}$ defines a self-dual
basis:
\begin{equation}
\tr  (\mu^i \mu^j)  = \delta_{ij},
\qquad i,j \in {3,5,6} .
\end{equation}
Here the trace of an element $\theta$ in this field
is defined as
\begin{equation}
\tr \theta = \theta + \theta^2 + \theta^4 .
\end{equation}

Following Refs.~\onlinecite{Wootters4} and \onlinecite{Gibbons},
we subsequently associate the self-dual basis elements, in
a two-dimensional phase-space coordinate representation,
with the two groups of three operators, each one
representing an ``axis'' in phase space:
\begin{eqnarray}
(\mu^3,0) & \leftrightarrow & \opx \identity \identity ,
\nonumber \\
(\mu^5,0) & \leftrightarrow & \identity \opx \identity , \\
(\mu^6,0) & \leftrightarrow & \identity \identity \opx ,
\nonumber
\end{eqnarray}
and
\begin{eqnarray}
(0,\mu^3) & \leftrightarrow & \opz \identity \identity ,
\nonumber \\
(0,\mu^5) & \leftrightarrow & \identity \opz \identity , \\
(0,\mu^6) & \leftrightarrow & \identity \identity \opz .
\nonumber
\end{eqnarray}
We map the eight-dimensional Hilbert space onto a $8 \times 8$
discrete phase space, and label the $x$ (horizontal) and $z$ axes
(vertical) with the field element sequence $\{0,1,\mu, \mu^2, \ldots,
\mu^6\}$. This gives us a phase-space coordinate in terms of
field-element powers: each phase-space point has a unique coordinate,
such as $(\mu^3, \mu^6)$, as can be seen in Fig.~\ref{Phase-space 1}.

\begin{figure}[b]
\begin{tabular}{c|c|c|c|c|c|c|c|c|}
\hline $\mu^6$ & $\; 2 \:$ & $\; 6 \:$& $\; 7 \:$& $\; 8 \:$& $\; 4 \:$& $\; 5 \:$& $\; 3 \:$& $\; 9\:$\\
\hline $\mu^5$ &2 & 7 & 8 & 9 & 5 & 6 & 4 & 3\\
\hline $\mu^4$ &2 & 8 & 9 & 3 & 6 & 7 & 5 & 4\\
\hline $\mu^3$ &2 & 5 & 6 & 7 & 3 & 4 & 9 & 8\\
\hline $\mu^2$ &2 & 4 & 5 & 6 & 9 & 3 & 8 & 7\\
\hline $\mu$   &2 & 9 & 3 & 4 & 7 & 8 & 6 & 5\\
\hline $1$     &2 & 3 & 4 & 5 & 8 & 9 & 7 & 6\\
\hline $0$     &o & 1 & 1 & 1 & 1 & 1 & 1 & 1\\
\hline         &$0$&$1$&$\mu$&$\mu^2$&$\mu^3$&$\mu^4$&$\mu^5$&$\mu^6$\\
\end{tabular}
\caption{The striation-generating curves corresponding to the MUB
construction defined by Table 1.} \label{Phase-space 1}
\end{figure}

One can now construct lines and striations (sets of eight parallel
lines, i.e., with no point in common) based on the algorithm for
the MUBs. The striations are uniquely defined by (any) one line
and the requirement of translational covariance. We shall,
for definiteness, take the striation-defining line as the
one that includes the origin, that is, a ray. Subsequently,
the seven other lines belonging to each striation can be
generated by translating the ray in a manner that will be
described below. In order to find a correspondence between the
phase-space lines, the translations, and MUBs, we first define
the unitary qubit-flip operators corresponding to the two sets
of generating operators (that is, the set of operators in a row
in the respective defining table). The unitary operator
$\exp(i \pi \opx /2) \otimes \identity \otimes \identity =
\hat{U}_x\otimes \identity \otimes \identity$ (with $\hat{U}_x =
i \hat{\sigma}_x$) rotates the first spin qubit around the
$x$ axis by the angle $\pi$. That is, the operator is an
eigenoperator of the set $\{ \opx \identity \identity,
\identity \opx \identity, \identity \identity \opx\}$, and flips the
first qubit of any eigenstate of $\{\opz \identity \identity,
\identity \opz \identity, \identity \identity \opz\}$ into the
orthogonal qubit state. In the same manner, local unitary flip
operators corresponding to the remaining five basis-defining
operators $\{\identity \opx \identity, \identity \identity \opx,\opz
\identity \identity, \identity \opz \identity, \identity \identity
\opz\}$ can be constructed.

In the phase-space representation these flip operators are
represented by a translation by the corresponding field
element ``vectors'', e.g., $\hat{U}_x\otimes \identity \otimes
\identity \leftrightarrow $ translation by addition of $(\mu^3,0)$,
and $\identity \otimes \hat{U}_z \otimes \identity \leftrightarrow$
translation by addition of $(0, \mu^5)$. The ray associated with the
set $\{ \opx \identity \identity, \identity \opx \identity,
\identity \identity \opx\}$ is hence generated by translation of the
origin $\{0,0\}$ by all additive (modulo~2) combinations of
$(\mu^3,0)$, $(\mu^5,0)$, and $(\mu^6,0)$. It is clear that this
will result in the horizontal line (ray) of Fig.~\ref{Phase-space
1}. (The origin is marked with an o because it belongs to all
striation-generating rays.) It is also evident that if every point
on this ray is transformed by any combination of the three
translations, the ray will remain invariant. This reflects the fact
that the state associated with the ray is an eigenvector of the
corresponding unitary qubit flip operators.

The rest of the lines of this striation can be obtained by
translation by all the combinations of the three vectors
$(0,\mu^3)$, $(0,\mu^5)$, and $(0,\mu^6)$ (respecting the
modulo~2 arithmetic and the field algebra). This results
in a set of eight parallel lines -- the striation
corresponding to the basis generated by the operators
$\{ \opx \identity \identity, \identity \opx \identity,
\identity \identity \opx\}$. In the same manner, one can
construct the striation (consisting of straight vertical
lines) representing the MUB generated by $\{ \opz \identity
\identity, \identity \opz \identity, \identity \identity \opz \}$.
The ray belonging to this set is labeled with  2s in
Fig.~\ref{Phase-space 1}.

To generate the third striation-generating ``ray'', we
follow the recipe given above~\cite{Klimov} and simply
multiply the first two rows of Table~\ref{table1} columnwise.
In doing so, we ignore overall phase factors. We get the
third set of MUB-generating operators $\{ \opy \identity \identity,
\identity \opx \opz, \identity \opz \opx\}$. These operators
correspond to the phase-space operators $(\mu^3,\mu^3)$,
$(\mu^5,\mu^6)$, and $(\mu^6,\mu^5)$. The ``ray'' generated
by displacement of the origin by all the combinations
of these translations is labeled with 3s in Fig.~\ref{Phase-space 1}.

The reason we have used the word ``ray'' within quotation
marks in the paragraph above is that ``ray'' 3 is in fact
a curve in phase-space. If a point belonging to it is denoted
$(\alpha,\beta)$, then the curve is parametrically
defined as
\begin{equation}
\alpha = \mu^3 \kappa + \mu^5 \kappa^2 + \mu^6 \kappa^4,
\qquad
\beta = \mu^2 \kappa + \kappa^2 + \mu^4 \kappa^4 ,
\end{equation}
where $\kappa$ is a parameter running over the field elements.
This parametrization corresponds to the explicit expression
\begin{equation}
\beta^2 + \mu \beta = \alpha^2 + \mu \alpha ,
\end{equation}
or, equivalently,
\begin{equation}
\beta =  \mu^6 \alpha + \mu^5 \alpha^4 + \mu^3 \alpha^2 ,
\end{equation} from which it is clear that $\alpha$ and $\beta$ are
not linearly related. In the following we will refer
to a curve passing through the origin as a homogeneous
curve.

From the table we see that, as required by a set of MUBs, each curve
crosses the curves of the other striations only once, at the origin
in Fig.~\ref{Phase-space 1} by our choice of depicting only
homogeneous curves. Every full striation can be generated from the
homogeneous curve by translation in the horizontal direction by
adding, in sequence, all the combinations of $(\mu^3,0)$,
$(\mu^5,0)$, and $(\mu^6,0)$. In physical space, this corresponds to
all possible combinations of spin flips around the $x$ axes. The
striations can equally well be generated by vertical translation by
all the combinations of $(0,\mu^3)$, $(0,\mu^5)$, and $(0,\mu^6)$,
corresponding to spin flips around the $z$ axes. In fact, we can
generate any striation from the corresponding homogeneous curve by
the translations corresponding to any of the other curve-generating
operators. For example, the operators $(\mu^3,\mu^3+\mu^5)$,
$(\mu^5,\mu^3)$, and $(\mu^6,\mu^6)$, corresponding to the first
three columns of the bottom row of Table~\ref{table1} will also
generate the same striation. This follows from the fact that the
relation between the bases is the same for any two of them. The
fully separable bases are not particular in this respect.

To obtain a Wigner function from the striations one more step
is  needed, namely, to associate each line in phase space with
a state. Again this involves an arbitrary choice that does not
lead to anything fundamentally new. As described in
Ref.~\onlinecite{Gibbons}, we can associate any basis eigenstate
to any line in a striation, but once this choice is made, the state
associated to a line obtained  by a particular translation must
correspond to the state obtained from the first by the corresponding
spin-flip transformation. For example, if we associate the vertical
ray in Fig.~\ref{Phase-space 1} (labeled with 2s) with the state
$|\uparrow \uparrow \uparrow \rangle_z$ (in a spin-1/2 representation
with $z$ as our spin axis), then the vertical line that includes the
point $(\mu^3+\mu^6,0)=(\mu^4,0) \leftrightarrow \hat{U}_x \otimes
\identity \otimes \hat{U}_x$ represents the state $\hat{U}_x \otimes
\identity \otimes \hat{U}_x | \uparrow \uparrow \uparrow \rangle_z =
| \downarrow \uparrow \downarrow \rangle_z$. As there are $N+1$
striations, and for each striation we can associate the generating
ray with any of $N$ state vectors, there are $N^{N+1}$
\textit{quantum nets} associated with this choice of phase-space
bases (single qubit rotations around the $x$ and $z$ axes). Of
these, one can group the possibilities in $N^{N-1}$
\textit{equivalence classes} each containing $N^2$
elements~\cite{Gibbons}. The elements are simply the $N \times N$
possible discrete translations of the space using all combinations
of the horizontal and vertical translation operators. Nevertheless,
we stress that in our approach, in contrast to that of
Wootters~\cite{Wootters4,Gibbons}, we associate curves in phase
space (and not merely lines) to states.

With the prerequisites above, for any phase-space point
$(\alpha,\beta)$ we can define Hermitian phase-space
point-operators $\hat{A}_{(\alpha,\beta)}$ as
\begin{equation}
\hat{A}_{(\alpha,\beta)}
= \sum_{k=1}^{9} \hat{\rho}_{k,(\alpha,\beta)} - \identity ,
\end{equation}
where $\hat{\rho}_{k,(\alpha,\beta)}$ is the density matrix
associated with the curve belonging to striation $k$ passing
through $(\alpha,\beta)$. From the mutual unbiasedness of the
orthonormal bases, it follows that
\begin{eqnarray}
\Tr  [ \hat{A}_{(\alpha,\beta)}] & = & 1 ,
\nonumber \\
& &  \\
\Tr [ \hat{A}_{(\alpha,\beta)}
\hat{A}_{(\alpha^\prime,\beta^\prime)} ] & = & 8
\delta_{\alpha,\alpha^\prime} \delta_{\beta,\beta^\prime} ,
\nonumber
\end{eqnarray}
where Tr (with upper case) denotes the ordinary Hilbert-space
trace operation. The discrete Wigner function $W(\alpha,\beta)$
of the density matrix $\hat{\rho}$ is then defined as
\begin{equation}
W (\alpha,\beta) = \frac{1}{8} \Tr
[ \hat{\rho} \hat{A}_{(\alpha,\beta)} ].
\end{equation}
This leads to the relations
\begin{eqnarray}
& \displaystyle
\sum_{\alpha,\beta} W(\alpha,\beta) =  1 , & \nonumber \\
&
\displaystyle
\hat{\rho}  =   \sum_{\alpha,\beta}
W(\alpha,\beta) \hat{A}_{(\alpha,\beta)} , & \\
& \displaystyle
\Tr (\hat{\rho}\hat{\rho}^\prime)  = 8 \sum_{\alpha,\beta}
W(\alpha,\beta) W^\prime(\alpha,\beta)  , & \nonumber
\end{eqnarray}
where
$W^\prime(\alpha,\beta)$ is the Wigner function of the
(arbitrary) density matrix $\hat{\rho}^\prime$. It is seen
that the phase-space point-operators, and consequently the
Wigner function, follow naturally from the construction
algorithm outlined above, based only on the generating
operator tables,  the irreducible polynomial (\ref{Eq: Irreducible}),
and the (arbitrary) association between basis states and
phase-space rays.

\begin{figure}[b]
\begin{tabular}{|c|c|c|c|c|c|c|c|}
\hline $\; 2 \:$ & $\; 6 \:$& $\; 5 \:$& $\; 8 \:$& $\; 3 \:$& $\; 7 \:$& $\; 4 \:$& $\; 9\:$\\
\hline 2 & 5 & 4 & 7 & 9 & 6 & 3 & 8\\
\hline 2 & 9 & 8 & 4 & 6 & 3 & 7 & 5\\
\hline 2 & 7 & 6 & 9 & 4 & 8 & 5 & 3\\
\hline 2 & 4 & 3 & 6 & 8 & 5 & 9 & 7\\
\hline 2 & 8 & 7 & 3 & 5 & 9 & 6 & 4\\
\hline 2 & 3 & 9 & 5 & 7 & 4 & 8 & 6\\
\hline o & 1 & 1 & 1 & 1 & 1 & 1 & 1\\
\hline
\end{tabular}
\hspace{1cm}
\begin{tabular}{|c|c|c|c|c|c|c|c|}
\hline $\; 2 \:$ & $\; 5 \:$& $\; 7 \:$& $\; 6 \:$& $\; 8 \:$& $\; 4 \:$& $\; 9 \:$& $\; 3\:$\\
\hline 2 & 7 & 9 & 8 & 3 & 6 & 4 & 5\\
\hline 2 & 8 & 3 & 9 & 4 & 7 & 5 & 6\\
\hline 2 & 6 & 8 & 7 & 9 & 3 & 1 & 2\\
\hline 2 & 9 & 4 & 3 & 5 & 8 & 6 & 7\\
\hline 2 & 4 & 6 & 5 & 7 & 3 & 8 & 9\\
\hline 2 & 3 & 5 & 4 & 6 & 9 & 7 & 8\\
\hline o & 1 & 1 & 1 & 1 & 1 & 1 & 1\\
\hline
\end{tabular}
\caption{The striation-generating lines corresponding to the MUB
construction defined by the second and fourth sets in Table 2.}
\label{Phase-space 3}
\end{figure}

The phase-space structure corresponding to the first set in
Table~\ref{table2} is obtained by making a mirror image of
Fig.~\ref{Phase-space 1} with respect to the diagonal through the
origin. As noted above, this stems from the fact that the
replacement $x \leftrightarrow z$ generates this set in
Table~\ref{table2} from the one in Table~\ref{table1}, and {\em vice
versa}. The phase-space structure corresponding to the other sets in
Table~\ref{table2} can be derived in the same manner. However, the
interrelation between the structures through qubit permutations is
not as easy to unveil in phase space as in physical space, as
demonstrated by Fig.~\ref{Phase-space 3}.

\section{The (3,0,6) MUB-structure Wigner function}
\label{The (3,0,6)}

As we have seen above, individual rotations of the
qubits around the $x$ and $z$ axes only generate one
MUB structure. To proceed further, one needs to
consider simultaneous rotations of two or more qubits.
The simplest one, in which we allow simultaneous rotation
of the first and the second qubits around the $x$  and $z$
axes, is shown in Table~\ref{table7}.

\begin{table}[h]
\caption{A table defining a (3,0,6) MUB structure.}
\begin{tabular}{ccc} \hline \hline
$\opz \opz \identity$ & $\identity \identity \opz$ &
$\opz \identity \identity$ \\
$\opx \opx \identity$ & $\identity \identity \opx$ &
$\opx \identity \identity$ \\  \hline \hline
\end{tabular}
\label{table7}
\end{table}

The rows of the table have the same eigenvectors as the
corresponding rows of Tables~\ref{table1} and \ref{table2} above.
Multiplying the rows columnwise, we obtain the operators  $\{ \opy
\opy \identity, \identity \identity \opy, \opy \identity
\identity\}$. These operators have the associated phase-space
translations $(\mu^3+\mu^5,\mu^3+\mu^5)$, $(\mu^6,\mu^6)$, and
$(\mu^3,\mu^3)$. It is evident that all the three operators make
``diagonal'' translations in phase space. It is also evident that
the corresponding MUB is triseparable: it is a $(3,0,6)$ structure.
The striation-generating rays of the phase-state structure are
displayed in Fig.~\ref{Phase-space 7}.

\begin{figure}[b]
\begin{tabular}{|c|c|c|c|c|c|c|c|}
\hline $\; 2 \:$ & $\; 8 \:$& $\; 6 \:$& $\; 4 \:$& $\; 9 \:$& $\; 7 \:$& $\; 5 \:$& $\; 3\:$\\
\hline 2 & 6 & 4 & 9 & 7 & 5 & 3 & 8\\
\hline 2 & 4 & 9 & 7 & 5 & 3 & 8 & 6\\
\hline 2 & 9 & 7 & 5 & 3 & 8 & 6 & 4\\
\hline 2 & 7 & 5 & 3 & 8 & 6 & 4 & 9\\
\hline 2 & 5 & 3 & 8 & 6 & 4 & 9 & 7\\
\hline 2 & 3 & 8 & 6 & 4 & 9 & 7 & 5\\
\hline o & 1 & 1 & 1 & 1 & 1 & 1 & 1\\
\hline
\end{tabular}
\hspace{1cm}
\begin{tabular}{|c|c|c|c|c|c|c|c|}
\hline $\; 2 \:$ & $\; 4 \:$& $\; 9 \:$& $\; 7 \:$& $\; 5 \:$& $\; 3 \:$& $\; 8 \:$& $\; 6\:$\\
\hline 2 & 7 & 5 & 3 & 8 & 6 & 4 & 9\\
\hline 2 & 8 & 6 & 4 & 9 & 7 & 5 & 3\\
\hline o & 1 & 1 & 1 & 1 & 1 & 1 & 1\\
\hline 2 & 6 & 4 & 9 & 7 & 5 & 3 & 8\\
\hline 2 & 3 & 8 & 6 & 4 & 9 & 7 & 5\\
\hline 2 & 5 & 3 & 8 & 6 & 4 & 9 & 7\\
\hline 2 & 9 & 7 & 5 & 3 & 8 & 6 & 4\\
\hline
\end{tabular}
\caption{The striation-generating rays corresponding to the
MUB construction defined by Table 3, left, and the lines resulting
from displacement of the rays by $(0,\mu^3)$, right.}
\label{Phase-space 7}
\end{figure}

This table has a particularly simple structure although it is built
using the same algorithm as all the preceding phase-space
structures. The rays in Fig.~\ref{Phase-space 7}, left side, are
really rays, and hence all striations will consist of lines. That
is, the different rays are generated by the equations $\beta =
\lambda \alpha$ and $\alpha = 0$, where the ``slope'' $\lambda$ for
each ray is a field element. Moreover, the rays are ``visually
straight''. This, e.g., means that Table~\ref{table7} generates
the same MUB under the interchange of axes $x
\leftrightarrow z$. We can once more generate six ``different''
tables, representing the same physical structure, by relabeling
the qubits. The simplicity of this figure is a chimera, however,
because only the rays are diagonal and ``visually straight''.
Translating the rays by, e.g., flipping the first qubit around
the $z$ axis (corresponding to a translation by $(0,\mu^3)$
yields the lines in the right part of Fig.~\ref{Phase-space 7}.
In this set of lines, one from each striation, only the
horizontal and the vertical lines remain ``visually straight''.

The interesting point is that there exist a physically different
$(3,0,6)$ MUB structure, also based on two single qubit-rotations
and one two-qubit rotation. It is shown in Table~\ref{table8}.

\begin{table}[t]
\caption{Another table defining a (3,0,6) MUB structure.}

\begin{tabular}{ccc} \hline \hline
$\opz \identity \identity$ & $\opz \opz \identity$ &
$\identity \identity \opz$ \\
$\opx \identity \identity$ & $\opx \opx \identity $ & $\identity
\identity \opx$ \\  \hline \hline
\end{tabular}
\label{table8}
\end{table}

\begin{figure}[t]
\begin{tabular}{|c|c|c|c|c|c|c|c|}
\hline $\; 2 \:$ & $\; 8 \:$& $\; 7 \:$& $\; 4 \:$& $\; 5 \:$& $\; 6 \:$& $\; 9 \:$& $\; 3\:$\\
\hline 2 & 9 & 8 & 5 & 6 & 7 & 3 & 4\\
\hline 2 & 5 & 4 & 8 & 9 & 3 & 6 & 7\\
\hline 2 & 6 & 5 & 9 & 3 & 4 & 7 & 8\\
\hline 2 & 7 & 6 & 3 & 4 & 5 & 8 & 9\\
\hline 2 & 4 & 3 & 7 & 8 & 9 & 5 & 6\\
\hline 2 & 3 & 9 & 6 & 7 & 8 & 4 & 5\\
\hline o & 1 & 1 & 1 & 1 & 1 & 1 & 1\\
\hline
\end{tabular}
\caption{Homogeneous curves corresponding to the
MUB construction defined by Table 4.} \label{Phase-space 9}
\end{figure}

The homogeneous curves corresponding to this structure are
displayed in Fig.~\ref{Phase-space 9}. A significant result
is that the structures represented by Tables~\ref{table7} and
\ref{table8} are the only physically inequivalent ones that
one can obtain from one double rotation and two single qubit
rotations.

\section{MUB structures (1,6,2) and (0,9,0)}
\label{Negative result}

It is known that two more MUB structures exist in the three-qubit
space, the (1,6,2) and (0,9,0)~\cite{Lawrence,Romero}. The first has
only one triseparable basis, whereas the second has none. It is
still possible to associate these with Wigner functions in a similar
manner, the only difference is that the corresponding Wigner
functions will have only one or no ``axes'' (striations composed of
lines). This means that most striations will consist of curves with
two, and sometimes four, points on the same horizontal or vertical
coordinate. To demonstrate this we use a $(1,6,2)$ structure, which
can be generated from the operator sets in Table~\ref{table9}.

\begin{table}[t]
\caption{A table defining a (1,6,2) MUB structure.}
\begin{tabular}{ccc} \hline \hline
$\identity \opz \identity$ & $\opz \opz \identity$ &
$\identity \identity \opz$ \\
$\opx \opy \identity$ & $\identity \identity \opx $ & $\opy \opz
\identity$ \\ \hline \hline
\end{tabular}
\label{table9}
\end{table}

\begin{figure}[t]
\begin{tabular}{|c|c|c|c|c|c|c|c|}
\hline $\; 2 \:$ & $\; 7 \:$& $\; 3 \:$& $\; 4 \:$& $\; 8 \:$& $\; 3 \:$& $\; 4 \:$& $\; 9\:$\\
\hline 2 & 1 & 7 & 1 & 9 & 4 & 8 & 4\\
\hline 2 & 9 & 6 & 6 & 6 & 8 & 7 & 6\\
\hline 2 & 5 & 1 & 9 & 5 & 7 & 1 & 8\\
\hline 2 & 3 & 8 & 7 & 1 & 1 & 9 & 3\\
\hline 2 & 4 & 4 & 8 & 7 & 9 & 5 & 5\\
\hline 2 & 8 & 9 & 5 & 3 & 5 & 3 & 7\\
\hline o & 6 & 5 & 3 & 4 & 6 & 6 & 1\\
\hline
\end{tabular}
\caption{Homogeneous curves corresponding to the MUB construction
defined by Table 5.}  \label{Phase-space 10}
\end{figure}

We see that the set involves ``simultaneous'' rotations of the first
and middle qubits around the $x$ and $z$ axes, effectively resulting
in a rotation around the $y$ axis. In phase space this means that by
choosing single qubit rotations around the $x$ and $z$ axes as our
``primitive'' operations, no set of striation-generating
translations decouple in horizontal and vertical translations.
Therefore, the correspondence with the translation and boost
operators $\hat{q}$ and $\hat{p}$, respectively, in the continuous
case is lost. The homogeneous curves corresponding to this structure
are depicted in Fig.~\ref{Phase-space 10}. We see that since there
is no triseparable basis involving only rotations around the $x$
axis, there is no horizontal ray. Instead, the curve labeled 1 can
parametrically be defined as
\begin{equation}
\alpha =\mu^2 \kappa^4 , \qquad
\beta = \mu^2 \kappa + \kappa^2 + \mu \kappa^4 ,
\end{equation}
which corresponds to the explicit expression
\begin{equation}
\beta^2 + \mu^4 \beta = \mu^3 \alpha^2 + \mu^2 \alpha ,
\end{equation}
or, in an equivalent form,
\begin{equation}
\beta =  \mu^3 \alpha + \mu^5 \alpha^2 + \mu^6 \alpha^4 .
\end{equation}
This in turn implies that the other curves have several points on
the same horizontal line. For example, curve 6 has four points on
the horizontal lines with $z$-coordinates 0 and $\mu^4$. This
reflects the fact that this is a curve invariant under the two
horizontal translations $(\mu^3+\mu^6,0)$ and $(\mu^5,0)$. By
applying the translation operators $(0,\mu^5)$, $(0,\mu^3+\mu^5)$,
$(0,\mu^6)$ one can subsequently generate striations from curves 1
and 3-9. However, this set of translation operators leave ray 2
invariant. To generate the whole striation from ray 2 we can, e.g.,
use the translation  set $(\mu^3+\mu^5,\mu^5)$, $(\mu^6,0)$, and
$(\mu^3,\mu^3+\mu^5)$, which are the generating translations of
curve 1.

To finally depict a ``severely disordered'' three-qubit Wigner
function (at least to the human eye), we use Table \ref{table10},
generating a (0,9,0) MUB, which gives rise to the nine striation
generating homogeneous curves depicted in Fig.~\ref{Phase-space 11}.
In spite of its disordered appearance, the Wigner function defined
from Table \ref{table10} inherits all the desirable features from
the continuous Wigner function except for (visually) straight axes
corresponding to $q$ and $p$. This follows from the fact that all
the MUB have entangled basis vectors.

\begin{table}[h]
\caption{A table defining a (0,9,0) MUB structure.}

\begin{tabular}{ccc} \hline \hline
$\opx \identity \identity$ & $\identity \opz \opy$ &
$\identity \opy \opz$ \\
$\opy \identity \identity$ & $\identity \opx \opz $ & $\identity
\opz \opx$ \\ \hline \hline
\end{tabular}
\label{table10}
\end{table}

\begin{figure}[h]
\begin{tabular}{|c|c|c|c|c|c|c|c|}
\hline $\; 9 \:$ & $\; 7 \:$& $\; 3 \:$& $\; 9 \:$& $\; 4 \:$& $\; 7 \:$& $\; 2 \:$& $\; 6\:$\\
\hline 4 & 5 & 3 & 6 & 8 & 8 & 5 & 2\\
\hline 7 & 6 & 3 & 2 & 5 & 4 & 7 & 5\\
\hline 3 & 7 & 8 & 6 & 2 & 7 & 8 & 4\\
\hline 6 & 8 & 3 & 8 & 9 & 2 & 9 & 4\\
\hline 3 & 6 & 2 & 1 & 4 & 1 & 1 & 1\\
\hline 3 & 2 & 5 & 5 & 9 & 4 & 9 & 6\\
\hline o & 1 & 1 & 9 & 1 & 5 & 7 & 8\\
\hline
\end{tabular}
\caption{Homogeneous lines corresponding to the MUB construction
defined by Table 6.}  \label{Phase-space 11}
\end{figure}

\section{Concluding remarks}
\label{Conclusions}

We have discussed the relation between MUBs having different
structures with the respect of the entanglement properties and
translationally covariant Wigner functions for three qubits. We have
shown that there exist three fundamentally different
constructions of the discrete Wigner functions if single qubit
rotations around two orthogonal axes are used as the generating
operations. To construct the Wigner function we need only a MUB
generating table and a field-element generating irreducible
polynomial. Other constructions can then be generated by qubit
permutation and by associating curves in the striations with
different states (the choice being arbitrary). The number
three is surprisingly small, and shows that requiring translational
covariance imposes severe restrictions.

We have also shown that for three-qubit MUBs that have only one or
none triseparable bases, the method based on the above-mentioned
generating operations still works. Here we do not know how many
fundamentally different constructions exist. However, these MUBs
are perfectly legitimate from a physical point of view,
although they have a more complex appearance than the Wigner
functions based on (2,3,4) and (3,0,6) MUBs.

We have not addressed here the interesting question of the
factorizability of the Wigner function, that has been
previously considered by Durt~\cite{Durt3} and Pittenger and
Rubin~\cite{Pittenger2}. This is certainly relevant in the framework
of quantum tomography, particularly from an experimental
viewpoint. In principle, it would suffice that one
phase-space point is trifactorizable in order that the other
operators are, since the translations themselves are factorizable in
virtue of translational covariance. Unfortunately, all our efforts
to determine if there exist some triseparable phase-space point
operator have failed. To the best of our knowledge, the problem
thus still remains open.

In general, the curves that define a MUB on $\mathbb{F}_{2^3}$ form
one-dimensional Abelian structures, which can be conveniently
parametrized as
\begin{equation}
\alpha (\kappa) = \nu_1 \kappa + \nu_2 \kappa^2 + \nu_3 \kappa^4 ,
\qquad
\beta (\kappa )= \eta_1 \kappa + \eta_2 \kappa^2 + \eta_3 \kappa^4 ,
\end{equation}
where $\kappa $ is a parameter and the coefficients $\nu_j$ and
$\eta_j$ take values on $\mathbb{F}_{2^3}$, so that
\begin{eqnarray}
\alpha (\kappa + \kappa^\prime ) & = & \alpha (\kappa ) +
\alpha (\kappa^\prime),  \nonumber \\
& & \\
\beta ( \kappa + \kappa^\prime ) & = & \beta ( \kappa ) +
\beta (\kappa^\prime ). \nonumber
\end{eqnarray}
 Hence, such curves are the simplest generalization of
``straight" Abelian structures (rays) of the form
\begin{equation}
\alpha =0, \qquad
\text{or}
\qquad
\beta =\lambda \alpha ,
\end{equation}
or equivalently
\begin{equation}
\alpha ( \kappa ) = \eta \kappa ,
\qquad
\beta ( \kappa ) = \zeta \kappa ,
\end{equation}
where $\eta$ and $\zeta $ are fixed field elements.

We finally observe that the structures studied by
Wootters~\cite{Wootters4,Gibbons} (and also by
Bandhyopadhyay \emph{et al}~\cite{Bandyopadhyay}
and Durt~\cite{Durt2}) always assume two
trifactorizable bases, so that only the
cases (2,3,4) and (3,0,6) are possible. The
existence of a third triseparable basis
depends on the choice for the field basis:
in the self-dual used in this paper, this is
always the case [so we are lead automatically
to the (3,0,6) structure], while other choices
bring the (2,3,4). In this respect, it is interesting
to note that the (3,0,6) is the only three-qubit MUB
phase-space structure (depicted in Fig.~\ref{Phase-space 7})
consisting solely of straight lines.

\acknowledgments{This work was supported by the Swedish
Foundation for International Cooperation in Research and
Higher Education (STINT), the Swedish Foundation for
Strategic Research (SSF), the Swedish Research Council
(VR), the Mexican CONACyT under grant 45704, and the
Spanish Research Project FIS2005-06714}.

\end{document}